\documentclass[preprint,12pt]{elsarticle}

 \usepackage{graphicx}

\usepackage{amssymb}
\usepackage{amsmath}

\journal{Physics Letters B}

\begin{document}

\begin{frontmatter}

\title{First measurement of kaonic helium-3 X-rays}

\author[LNF]{M.~Bazzi}
\author[UVCA]{G.~Beer}
\author[POLI]{L.~Bombelli}
\author[LNF,IFIN]{A.M.~Bragadireanu}
\author[SMI]{M.~Cargnelli}
\author[LNF]{G.~Corradi}
\author[LNF]{C.~ Curceanu~(Petrascu)}
\author[LNF]{A.~d'Uffizi}
\author[POLI]{C.~Fiorini}
\author[POLI]{T.~Frizzi}
\author[ROMA]{F.~Ghio}
\author[ROMA]{B.~Girolami}
\author[LNF]{C.~Guaraldo}
\author[TKY] {R.S.~Hayano}
\author[LNF,IFIN]{M.~Iliescu}
\author[SMI]{T. Ishiwatari\corref{cor1}}
\ead{tomoichi.ishiwatari@assoc.oeaw.ac.at}
\cortext[cor1]{T. Ishiwatari}
\author[RIKEN]{M.~Iwasaki}
\author[SMI,MUN]{P.~Kienle}
\author[LNF]{P.~Levi~Sandri}
\author[POLI]{A.~Longoni}
\author[SMI]{J.~Marton}
\author[LNF]{S.~Okada}
\author[LNF]{D.~Pietreanu}
\author[IFIN]{T.~Ponta}
\author[LNF]{A.~Rizzo}
\author[LNF]{A.~Romero Vidal}
\author[LNF]{A.~Scordo}
\author[TKY] {H. Shi}
\author[LNF,IFIN]{D.L.~Sirghi}
\author[LNF,IFIN]{F.~Sirghi}
\author[TKY]{H.~Tatsuno}
\author[IFIN]{A.~Tudorache}
\author[IFIN]{V.~Tudorache}
\author[LNF]{O.~Vazquez~Doce}
\author[SMI]{E.~Widmann}
\author[SMI]{B.~W\"{u}nschek}
\author[SMI]{J.~Zmeskal}
\author[]{(SIDDHARTA collaboration)}

\address[LNF]{INFN, Laboratori Nazionali di Frascati, Frascati (Roma), Italy}
\address[UVCA]{Dep. of Phys. and Astro., Univ. of Victoria, Victoria B.C., 
Canada}
\address[POLI]{Politechno di Milano, Sez. di Elettronica, Milano, Italy}
\address[IFIN]{IFIN-HH, Magurele, Bucharest, Romania}
\address[SMI]{Stefan-Meyer-Institut f\"{u}r subatomare Physik, 
Vienna, Austria}
\address[ROMA]{INFN Sez. di Roma I and Inst. Superiore di Sanita, Roma, Italy}
\address[TKY]{Univ. of Tokyo, Tokyo, Japan}
\address[RIKEN]{RIKEN, The Inst. of Phys. and Chem. Research, Saitama, Japan}
\address[MUN]{Tech. Univ. M\"{u}nchen, Physik Dep., Garching, Germany}

\begin{abstract}
The first observation of the kaonic $^3$He $3d \to 2p$ transition
was made, using slow $K^-$ mesons stopped in a gaseous $^3$He target. 
The kaonic atom X-rays were detected with large-area 
silicon drift detectors  using the timing information 
of the $K^+K^-$ pairs of $\phi$-meson decays produced by the 
DA$\Phi$NE $e^+e^-$ collider. The strong interaction
shift of the kaonic $^3$He $2p$ state was determined to be 
$ -2 \pm 2 \mbox{ (stat)} \pm 4 \mbox{ (syst)} \mbox{ eV}$.
\end{abstract}


\end{frontmatter}


\section{Introduction}
\label{intro}

Low-lying energy levels of kaonic atoms are shifted and broadened 
due to the strong interaction between the antikaon and nucleus. 
These shifts and widths are fundamental data for studies of the low-energy $\bar{K}N$ 
interaction, and generally of  low-energy QCD in the strangeness sector.
Although kaonic atom X-ray data have been taken using many target 
materials, little was known about kaonic atoms with $Z=1$ and $2$.
Concerning kaonic helium, no data was available for kaonic $^3$He, while, until recently,
for kaonic $^4$He there was a discrepancy between experimental results 
and theoretical predictions \cite{khe1,khe2,khe3,puzzle}.
This discrepancy was eventually solved in recent years by the E570 and 
SIDDHARTA experiments \cite{E570,sidd-khe}, where reliable experimental data were obtained.
As soon as a low-energy kaon beam, allowing the use of gaseous target, and fast-timing X-ray detectors
became available, for the first time, kaonic $^3$He X-rays 
could be measured by the SIDDHARTA experiment.

The SIDDHARTA experiment uses low-energy kaons generated by the DA$\Phi$NE collider,
which may be stopped efficiently in a gaseous target in a small volume.
Large-area silicon drift detectors (SDDs) -- developed in the framework of 
the SIDDHARTA collaboration -- were used as X-ray detectors 
for the first time at a collider machine.
Using these SDDs, it was possible to determine X-ray energies with a precision of a few eV in the
energy range up to about 10 keV, based on high suppression of background events.

%

In this paper, we report on this first measurement of the kaonic $^3$He X-rays, which is
extremely important in view of a possible shift of the kaonic $^3$He $2p$ level,
as well as a possible isotope difference of the $2p$ level shifts between kaonic $^3$He and $^4$He
depending on the strength of the $K^-$-$^3$He and $K^-$-$^4$He interaction \cite{akaishi}.

\section{The SIDDHARTA experimental setup}
\label{exp}
\begin{figure}[htbp]
\begin{center}
\centerline{\includegraphics[width=8cm]{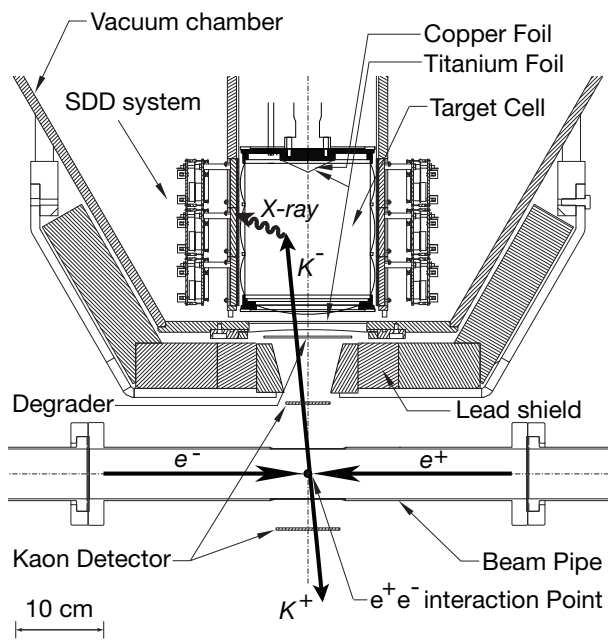}}
\caption{An overview of the experimental setup. The whole 
system was installed at the interaction point of DA$\Phi$NE.}
\label{fig1}
\end{center}
\end{figure}

The SIDDHARTA setup was installed at the $e^+e^-$ interaction point of the DA$\Phi$NE collider.
It consists of an X-ray detection system, a cryogenic target system, 
and a kaon detector, as shown in Fig. \ref{fig1}.  

Helium-3 gas at a temperature of 20 K and a pressure of 1 bar was used 
as a target. The gas was contained in a cylindrical target cell
(with a radius of 72 mm, and a height of 155 mm), made of 
75-$\mu$m thick Kapton foils.

Large area silicon-drift detectors (SDDs) having an active area 
of 1 cm$^2$ each and a thickness of 450 $\mu$m \cite{sdd1,sdd,NIM-tomo} were used 
for X-ray detection. A total active area of 144 cm$^2$ was installed 
with a distance of 78 mm between the SDDs and the target central axis. 
The SDDs were cooled to a temperature of 170 K with a stability of $\pm 0.5$ K.

The positions at which the SDDs were installed differed from those in the setup 
used for kaonic $^4$He in \cite{sidd-khe}. Together with a larger size of the target, 
the acceptance of the SDDs was improved by a factor of about 2.6.

$K^+K^-$  pairs produced by  $\phi$ decay were detected by two scintillators installed  
above and below the beam pipe at the interaction point (called ``the kaon detector''). 
The scintillator installed below the beam pipe has a size of 72 $\times$ 72 mm$^2$ and a thickness of 1.5 mm,
while the one installed above the pipe has a smaller size of 49 $\times$ 45 mm$^2$  
and a thickness of 1.5 mm. 
Above the upper scintillator a degrader was installed to degrade the kaon energy 
so that the $K^-$ mesons are stopped in the $^3$He target volume. 

High intensity X-ray lines for energy calibration
were periodically provided by irradiating
thin foils of titanium and copper with 
an X-ray tube to excite them. They were installed
at the interaction point,
replacing the kaon detector. 

Two types of data were taken with the $e^+ e^-$ beams.
The first type (``production'' data) is data taken with the kaon detector and degrader, 
to be used for collection of kaonic atom X-ray events.
The second type (``X-ray tube'' data) is data taken with the X-ray tube and the Ti and Cu foils.
These X-ray tube data were taken periodically (typically every several hours),
to be used for the determination of the energy scale of each SDD, and for monitoring
temporal changes in the positions of the Ti and Cu X-ray peaks.


Energy data of all the X-ray signals detected by the SDDs were recorded using a 
specially designed data acquisition system. Time differences between the X-ray signals in the SDDs
and the coincidence signals in the kaon detector  were recorded using clock signals with a 
frequency of 120 MHz, whenever the X-ray signals occurred within
 a time window of 6 $\mu$s. In addition, time differences between the coincidence signals 
in the kaon detector and the clock pulses delivered by DA$\Phi$NE were recorded.

The kaonic $^3$He X-ray data were taken for about 4 days in November of 2009.
In this period, an integrated luminosity of 17 pb$^{-1}$ was collected, which corresponds 
to about $2 \cdot 10^6$ kaons detected by the kaon detector.

\section{Analysis of kaonic helium X-ray data}

\begin{figure}[htbp]
\begin{center}
\centerline{\includegraphics[width=12cm]{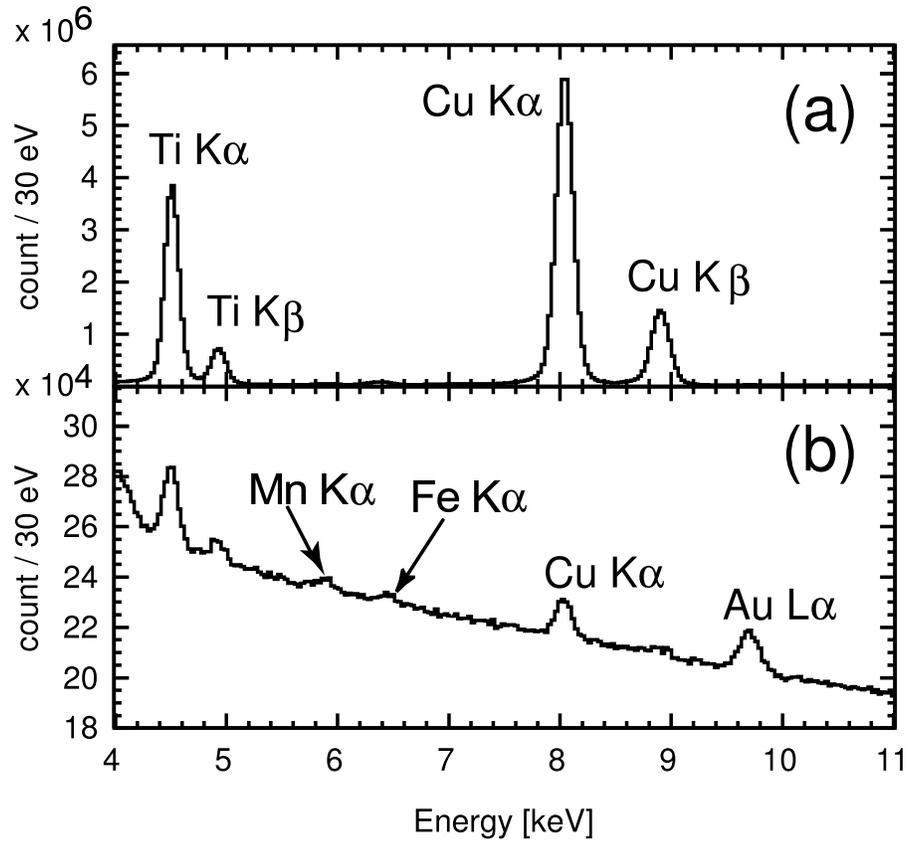}}
\caption{X-ray energy spectra of the SDDs, where data of all the selected SDDs were summed:
(a) data taken with the X-ray tube, and
(b) data uncorrelated to the kaon production timing in the production data.
The peak positions of the Ti, Cu, and Au fluorescence X-ray lines in Fig. (b) were used 
to determine the accuracy of the energy scale.}
\label{fig3}
\end{center}
\end{figure}

First, the X-ray tube data were analyzed. Energy spectra of each SDD 
contain Ti and Cu K$\alpha$ peaks with high statistics, mainly 
induced by radiation from the X-ray tube. Since each SDD has a different gain, 
the energy scale was determined using the known X-ray energies of the Ti and Cu lines. 
In addition, SDDs having good performance were selected, based on 
energy resolution, peak shape, and stability during the measurements \cite{refnim}. 
The energy spectrum of the X-ray tube data is shown in Fig. \ref{fig3}(a), 
where data of all the selected SDDs were summed. More detailed information can 
be found in \cite{refnim}. 

The production data were then analyzed using the energy scale 
determined from the X-ray tube data after corrections for
temporal fluctuations of the peak positions. The production data are 
categorized as two types, based on whether or not the coincidence signals 
between the SDD and kaon detector occurred within 
a coincidence window of 6 $\mu$s. One type contains 
X-ray events correlated with the kaon coincidence 
(triple coincidence data), providing kaonic atom X-ray energy 
spectra with a high background suppression.
The other type contains X-ray events uncorrelated with the 
kaon coincidence (non-coincidence data), providing
large statistics of background events, as well as X-ray lines
from the target materials induced by the beam background.

Figure \ref{fig3}(b) shows the energy spectrum of the non-coincidence data.
The Ti and Cu peaks are seen at an energy of 4.5 keV and 8.0 keV, respectively. These peaks
were produced by the Ti and Cu foils installed on the top of the target,
as well as by the Ti foil, which is a part of the degrader.
In addition,  the Au L$\alpha$ line is seen at 9.7 keV, which was 
produced by the material of the SDD support structure.
The Mn K$\alpha$ and Fe K$\alpha$ peaks were seen at 5.9 keV and 6.4 keV,
which were produced by a $^{55}$Fe source and the Fe foil installed on the
top of the degrader during part of the measurement.

The accuracy of the energy scale in the production data 
was evaluated using the peak positions of the Ti, Cu, and Au lines. 
For the evaluation, the data taken with other target gases
in the same experimental setup were also used, to 
reduce the effect of statistical fluctuations in the peak positions.
In the fit of these lines, X-ray energies determined by high-resolution measurements
were used as reference values \cite{refti,refmn,refau}, 
where their natural linewidth and asymmetric components
were also taken into account.

As a result of the fit, the energy scale in the production data is 
shifted by $-6.5$ eV compared to the reference data.
The accuracy of the energy determination was $\pm 3.5$ eV 
in the energy region from 4 keV to 10 keV \cite{refnim}.

This shift is due to the instability of our electronics caused by 
differing conditions between the production and X-ray tube data. 
The largest contribution is a rate-dependent effect on the SDDs. 
The hit rate of  the SDDs in the X-ray tube data was about 10 times 
higher than that in the production data, causing the observed shift.  

The shift needs a correction to determine X-ray energies on an absolute energy scale.
A value on the absolute energy scale ($E_{\rm exp}$) is calculated
by adding the correction term ($ \varepsilon $) to a fit value
in the production data ($E_{\rm fit})$:
\begin{equation}
 E_{\rm{exp}} =  E_{\rm{fit}} + \varepsilon,
\end{equation}
where
\begin{equation}
\varepsilon = + 6.5 \pm 3.5 \mbox{ eV}
\label{peakshift}
\end{equation}
The uncertainty of $\varepsilon$ is used for the evaluation of  a systematic error
in the energy determination.

\begin{figure}[htbp]
\begin{center}
\centerline{\includegraphics[width=10cm]{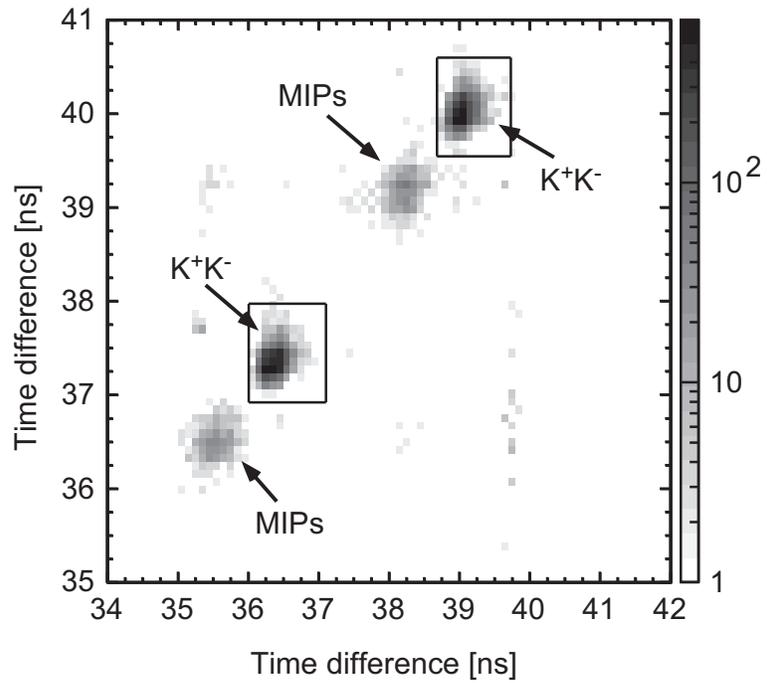}}
\caption{Timing spectrum of the two scintillators in the kaon detector. 
The time difference between the clock signals delivered by DA$\Phi$NE and the 
coincidence of the two scintillators is shown. The $K^+K^-$ and MIPs 
coincidence events are marked in the figure. 
The regions marked with a rectangle were accepted as  timing windows of 
the $K^+K^-$ coincidence.}
\label{fig2}
\end{center}
\end{figure}

The timing information of the coincidence data was analyzed to 
reject X-ray events uncorrelated to the $K^+K^-$ production timing.

The charged kaons were identified by means of the time-of-flight technique 
in the kaon detector \cite{sidd-khe,kmon}. A correlation of the time difference 
on the two scintillators is shown in Fig. \ref{fig2},
where the gray scale in the figure corresponds to the number of events per bin on a 
logarithmic scale. The events corresponding to $K^+K^-$ pairs were marked in the figure, as well as
the events of fast minimum-ionizing particles (MIPs) which passed through the two scintillators 
in coincidence. 
Because a half frequency of the beam synchronous timing signal was used for the start timing 
of the time-of-flight, the $K^+ K^-$ pair events appeared in two different timing regions 
in this timing spectrum \cite{refnim,kmon}.
The regions marked with a rectangle were accepted as a timing window for
the $K^+K^-$ coincidence.

\begin{figure}[htbp]
\begin{center}
\centerline{\includegraphics[width=10cm]{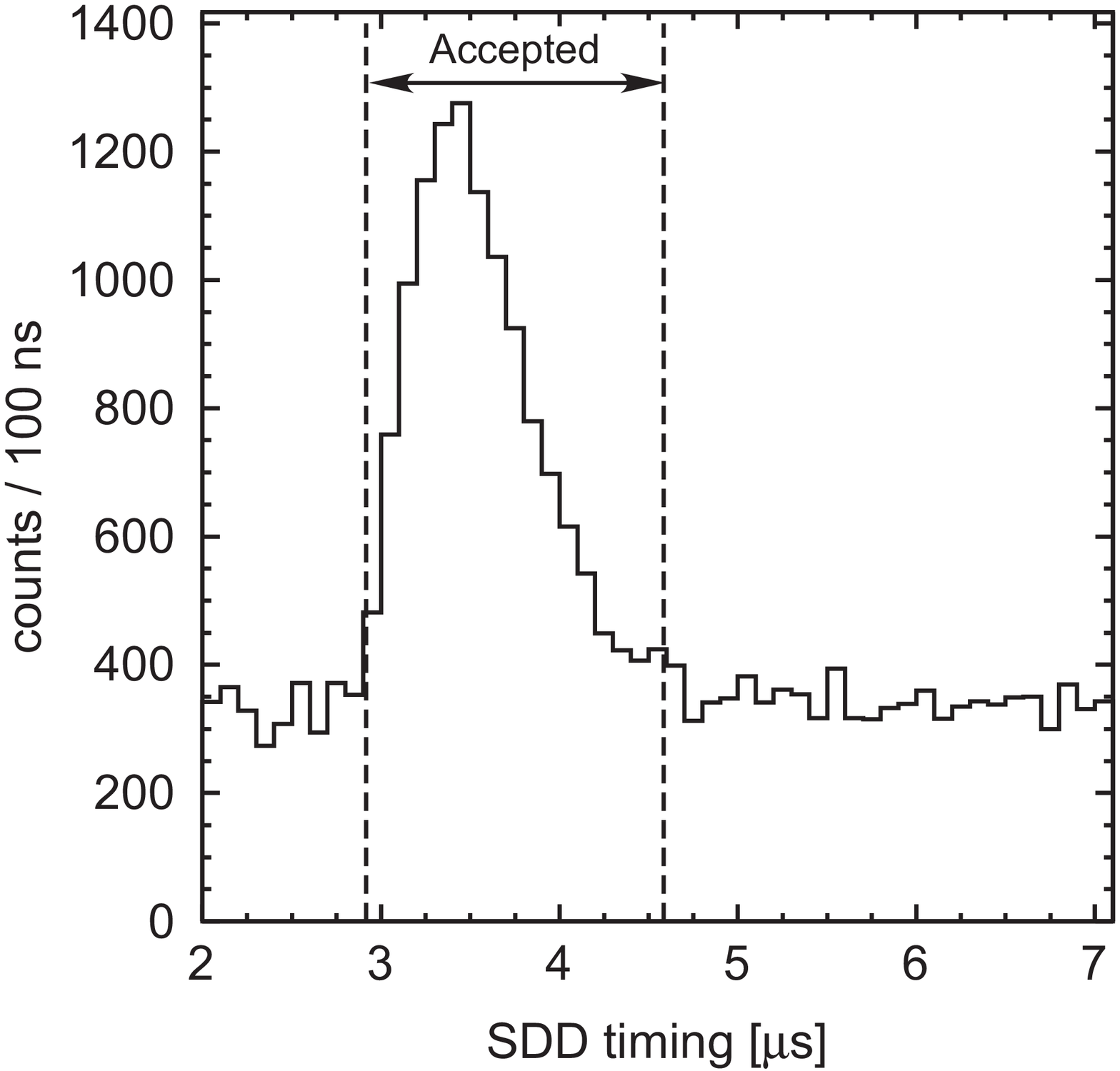}}
\caption{Time spectrum of the SDDs. The time difference between the
$K^+K^-$ coincidence and SDD X-ray hits was plotted. The peak region
corresponds to the coincidence of the $K^+K^-$ and X-ray events. 
A region from 2.9 $\mu$s to 4.6 $\mu$s was 
accepted as a timing window of the triple coincidences.}
\label{fig4}
\end{center}
\end{figure}

The time-difference spectrum of the $K^+K^-$ pair events in the kaon detector
and the X-ray events in the SDDs is shown in Fig. \ref{fig4}. The origin of the 
horizontal axis is arbitrary because of a delay time of the electronics. 
The peak in the figure corresponds to the coincidence events of the $K^+K^-$ pairs 
and X-rays (triple coincidences). A region from 2.9 $\mu$s to 4.6 $\mu$s was 
selected as the timing window of the coincidences.

\begin{figure}[htbp]
\begin{center}
\centerline{\includegraphics[width=10cm]{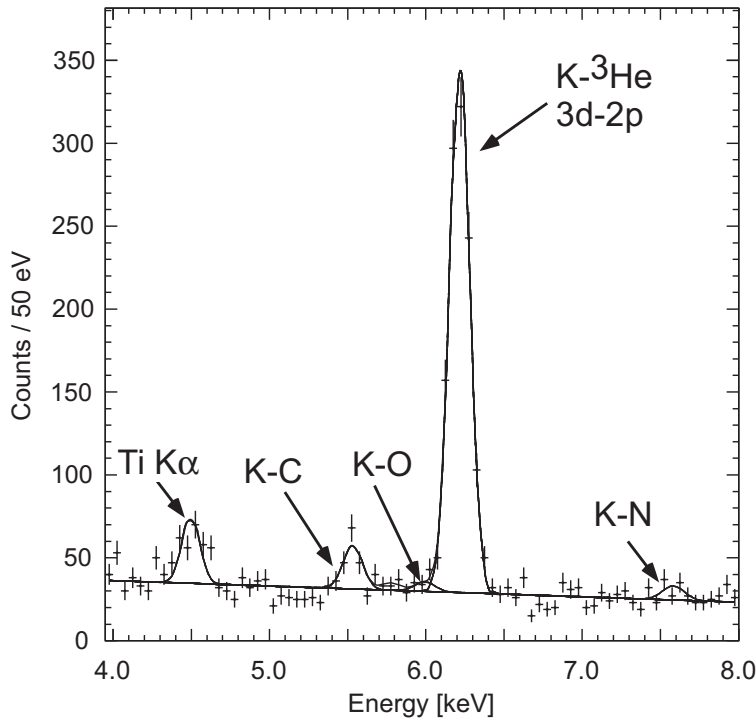}}
\caption{Energy spectrum of the kaonic $^3$He X-rays in coincidence 
with the $K^+K^-$ events. The kaonic $^3$He $3d \to 2p$ transition is
seen at 6.2 keV. Together with this peak, small peaks are seen, which
are the kaonic atom X-ray lines produced by kaons stopping in the target window
made of Kapton (Polyimide), and the Ti K$\alpha$ line at 4.5 keV.
}
\label{fig5}
\end{center}
\end{figure}

Figure \ref{fig5} shows the X-ray energy spectrum of the SDDs in the triple coincidence timing, 
where the energy scale determined from the X-ray tube data was used for the horizontal axis. 
A peak seen at 6.2 keV is identified as the kaonic $^3$He L$\alpha$ line 
(the $3d \to 2p$ transition). Along with this peak, other small peaks are seen. 
They are identified as the Ti K$\alpha$ line at 4.5 keV, 
the kaonic carbon $6h \to 5g$ transition at 5.5 keV, 
the kaonic oxygen $7i \to 6h$ at 6.0 keV, and  the kaonic nitrogen $6h \to 5g$  at 7.6 keV. 
The kaonic atom X-rays were produced by kaons stopping in the target window
made of Kapton (Polyimide) (${\rm C}_{22}{\rm H}_{10}{\rm O}_{5}{\rm N}_{2}$).
The identification of these lines was confirmed in the analysis of data taken 
with other target gases (hydrogen, deuterium, and helium-4).

The kaonic helium L$\alpha$ peak was fitted with a Voigt function, 
which is a convolution of a Gaussian and a Lorentzian.
The fit lines are shown in the figure. The energy resolution of the
peak is consistent with the value obtained from the non-coincidence data
(about 150 eV (FWHM) at 6.2 keV).
The peak position was determined to be 
\begin{equation}
E_{{\rm fit}} = 6216.5 \pm 2.4 \mbox{ eV}.
\end{equation}
Using the correction term $\varepsilon$ in Eq. (\ref{peakshift}), the 
absolute energy of the kaonic $^3$He $3d \to 2p$ transition was then determined to be:
\begin{eqnarray}
E_{{\rm exp}} &=& E_{{\rm fit}} + \varepsilon \\ \nonumber
              &=& 6223.0 \pm 2.4 \mbox{ (stat)} \pm 3.5 \mbox{ (syst)} \mbox{ eV},
\end{eqnarray}
where the second term is the statistical error, and the third term is the systematic error.
The latter was evaluated from the accuracy of the energy determination ($\pm$ 3.5 eV).
Other contributions to the systematic error 
(e.g. effects of timing region selection and contributions of
the kaonic oxygen line at 6.0 keV) are negligible.


\section{Discussion and conclusions}
The energy of the kaonic $^3$He $3d \to 2p$ transition ($E_{{\rm e.m.}}$)
was calculated using the Klein-Gordon equation together with an energy shift caused by 
the vacuum polarization effect. The formula given in \cite{Uehling-pionic} was
used for the calculation of the vacuum polarization effect, where the first order
of the Uehling potential was taken into account. The energy levels of 
the $3d$ and $2p$ states are tabulated in Table \ref{tab:calc}. 
The calculated energy ($E_{{\rm e.m.}}$) of the kaonic $^3$He $3d \to 2p$ transition 
was:
\begin{equation}
E_{{\rm e.m.}} = 6224.6 \mbox{ eV.}
\end{equation}
The contribution from the higher-order vacuum polarization terms
is estimated to be 0.2 eV, and  the uncertainty due to the kaon mass error 
is about $\pm 0.2$ eV. Other corrections (e.g. a recoil effect, a charge-radius effect) 
are negligibly small, compared to the above terms.

\begin{table}[ht]
\centering
\begin{center}
\caption{Calculated energy levels of the kaonic $^3$He $3d$ and $2p$ states.
The calculations used the Klein-Gordon equation (K.G.), and 
the vacuum polarization effect (V.P.) from 
the first-order term of the Uehling potential.
In the last line, the energy of kaonic $^3$He $3d \to 2p$ transition is shown.
\label{tab:calc}
}
\centering
\begin{tabular}{crrr}
\hline \hline
Level& K.G. [eV] & V.P. [eV] & Total [eV] \\
\hline
$2p$   & -11179.6 &-15.4 & -11195.0 \\
$3d$   & -4968.6  &-1.9  & -4970.5  \\
\hline
\hline
$3d \to 2p$ & 6211.0 & 13.5 & 6224.6\\
\hline
\hline
\end{tabular}
\end{center}
\end{table}


The strong-interaction shift $\Delta E_{2p}$ of the kaonic $^3$He $2p$ state is
obtained from the difference between the experimentally determined value
$E_{{\rm exp}}$ and the QED calculated value $E_{{\rm e.m.}}$ 
(The strong-interaction shift of the $3d$ state is negligibly small).
The  result is:
\begin{eqnarray}
\Delta E_{2p} &=& E_{{\rm exp}} - E_{{\rm e.m.}}\nonumber \\
&=& -2 \pm 2 \mbox{ (stat)} \pm 4 \mbox{ (syst)} \mbox{ eV,}
\end{eqnarray}
where the second term denoted as (stat) is the statistical error and
the third term denoted as (syst) is the systematic error.

Using the same setup as well as the same measuring and analysis procedures, 
kaonic $^4$He $3d \to 2p$ X-rays were measured over short periods 
for a very first look at a possible isotope shift between kaonic $^3$He 
and $^4$He \cite{akaishi}.  The strong-interaction shift of the kaonic $^4$He $2p$ state
was determined to be 
$
\Delta E_{2p} = +5 \pm 3 \mbox{ (stat)} \pm 4 \mbox{ (syst)} \mbox{ eV.}
$
This result is in agreement, within the errors, with the results reported by
the E570 \cite{E570} and SIDDHARTA \cite{sidd-khe} collaborations.
Since the present results both of the kaonic $^3$He and $^4$He shifts
were determined with the same procedures,
their difference gives directly a first indication that the kaonic $^3$He-$^4$He 
isotope shift is rather small, which is expected also in theories \cite{akaishi,friedman}.

In conclusion, for the first time, the energy of the kaonic $^3$He $3d \to 2p$ transition was 
measured using a gaseous $^3$He target in the SIDDHARTA experiment. 
The strong-interaction shift of the kaonic $^3$He $2p$ state was determined to be 
$\Delta E_{2p} = -2 \pm 2 \mbox{ (stat)} \pm 4 \mbox { (syst)}$ eV.

\section*{Acknowledgement}
We thank C. Capoccia, B. Dulach, and D. Tagnani from
LNF-INFN; and H. Schneider, L. Stohwasser, and  D. St\"{u}ckler
from Stefan-Meyer-Institut,
 for their fundamental contribution in designing and building
the SIDDHARTA setup.
We thank as well the DA$\Phi$NE staff for the excellent working
conditions and permanent support.
Part of this work was supported by HadronPhysics I3 FP6 European
Community program, Contract No. RII3-CT-2004-506078; 
the European Community Research Infrastructure Integrating 
Activity ``Study of Strongly Interacting Matter'' 
(HadronPhysics2, Grant Agreement No. 227431) under
the Seventh Framework Programme of EU;
Austrian Federal Ministry of Science and Research BMBWK  
650962/0001 VI/2/2009; Romanian National Authority for Scientific Research, 
Contract No. 2-CeX 06-11-11/2006; Grant-in-Aid for Specially 
Promoted Research (20002003), MEXT, Japan; and the Austrian Science
Fund (FWF): [P20651-N20].

\end{document}